\documentclass[twocolumn,preprintnumbers,nofootinbib]{revtex4-1}
\usepackage{graphicx} 
\usepackage{latexsym} 
\usepackage{mathptm}
\usepackage{amsmath}
\usepackage{amssymb}
\usepackage{graphicx}

\def\beq{\begin{equation}}
\def\eeq{\end{equation}}
\def\beqn{\begin{eqnarray}}
\def\eeqn{\end{eqnarray}}

\newcommand{\intsum}{\hspace*{1mm}{\mbox{\Large$\Sigma$}}\hspace*{-3.8mm}\int}
 \newcommand{\quabla}{  \widehat{\stackrel{}{\Box}}}

\begin{document}
 
\title{Quantum superpositions of the speed of light}
\author{Sabine Hossenfelder \thanks{hossi@nordita.org}}
\affiliation{Nordita, Roslagstullsbacken 23, 106 91 Stockholm, Sweden}

\begin{abstract}
While it has often been proposed that, fundamentally, Lorentz-invariance is not respected in
a quantum theory of gravity, it has been difficult to reconcile deviations from Lorentz-invariance
with quantum field theory. The most commonly used mechanisms either break Lorentz-invariance
explicitly or deform it at high energies. However, the former option is very tightly constrained 
by experiment already, the latter generically leads to problems with locality. We show here
that there exists a third way to integrate deviations from Lorentz-invariance
into quantum field theory that circumvents the problems of the other approaches. The way this
is achieved is an extension of the standard model in which photons can have different speeds
without singling out a preferred restframe, but only as long as they are in a
quantum superposition. Once a measurement has been made, observables are subject 
to the laws of special relativity, and the process of measurement introduces a preferred
frame. The speed of light can take on
different values, both superluminal and subluminal (with respect to the usual value of the
speed of light), without the need for Lorentz-invariance violating operators and without 
tachyons. 
We briefly discuss the relation to deformations of special relativity and 
phenomenological consequences.

\end{abstract}
\maketitle

\section{Introduction}

The speed of light plays an important role for the Lorentz-group and the physics of
special relativity (SR). It is the only speed that remains
invariant under a change of reference frame, and it determines the causal structure of space-time. 
Most importantly, the speed of light is the asymptotic limit
of the speed of accelerated massive bodies, and information cannot be transmitted any faster. 
The derivation of these properties from the symmetries of Minkowski-space is straight-forward
and SR has been experimentally confirmed to high precision. 
However, Lorentz-symmetry might not be respected by the yet-to-be found theory
of quantum gravity, and in fact deviations from Lorentz-symmetry are the so far most promising
route to make contact between theoretical approaches to quantize gravity and observation \cite{Mattingly:2005re}.

The maybe most obvious way that deviations from Lorentz-invariance can make themselves noticeable is a breaking
of Lorentz-invariance by the existence of a preferred frame. The preferred frame defines a timelike vector field,
and one expects this field to couple to other fields of the standard model (SM). 
Such a breaking of Lorentz-invariance in extensions of the SM is very
strongly constrained already \cite{Kostelecky:2008ts}. The introduction of a fundamental preferred frame also 
leaves open the question why, if not exact, Lorentz-symmetry is still approximately 
exact to high precision, in the sense that relevant operators and operators of dimension 5 that couple
to the timelike vector field are so strongly suppressed. Therefore, Lorentz-invariance violating
operators in the SM face both experimental and theoretical challenges.

An alternative to the introduction of a preferred frame is to make different values
of the speed of light compatible with observer-independence by modifying the action of the Poincar\'e-group. 
This enables the observer-invariance of an energy-dependent speed of light, and has been
developed in an approach known as ``Deformed Special Relativity'' ({\sc DSR})
\cite{AmelinoCamelia:2000ge,KowalskiGlikman:2001gp,AmelinoCamelia:2002wr,Magueijo:2002am}.
Such deformations of special relativity are intimately related to non-commutative geometry, 
an idea that dates back to Snyder in 1947 \cite{Snyder}, and that has received a lot of attention since
it was shown to arise by quantum deformations of Poincar\'e symmetry \cite{Majid:1994cy}. The
relation to {\sc DSR} was made in \cite{KowalskiGlikman:2003we} and has given rise to many
related works that have entered the literature under the names of modified
commutation relations, minimal length deformed quantum mechanics, or generalized uncertainty. 
These frameworks all explicitly or implicitly make use of deformations of special relativity.

{\sc DSR} has been motivated by Loop Quantum Gravity, though no rigorous derivation exists to
date. There are however non-rigorous arguments that {\sc DSR} may emerge from a semiclassical limit 
of quantum gravity theories in the form of an effective field theory with  an energy dependent metric \cite{AmelinoCamelia:2003xp},
or that {\sc DSR} (in form of the $\kappa$-Poincar\'e algebra) may result from a version of path integral 
quantization \cite{KowalskiGlikman:2008fj}. In addition it has been shown that
 in 2+1 dimensional gravity coupled to matter, the gravitational degrees of freedom can be integrated out, leaving 
an effective field theory for the matter which is a quantum field theory on $\kappa$-Minkowski space-time, realizing
a particular version of {\sc DSR} \cite{Freidel:2003sp}. Recently, it has also been suggested that
{\sc DSR} could arise via Loop Quantum Cosmology \cite{Bojowald:2009ey}. Originally formulated in momentum space, it has however proven difficult to extend the
formalism of {\sc DSR} to position space, and the so-far pursued attempts lead 
to macroscopic non-localities. The interpretation and relevance of these
non-localities is subject of an ongoing discussion. (More on this in section \ref{causality}). 

It has been proposed \cite{KowalskiGlikman:2006vx,Smolin:2010xa} that {\sc DSR} is a classical relic of the quantum gravitational
regime in the following sense. To modify the structure of momentum space and
the action of the Lorentz-group on it, one needs a constant of dimension
mass that one can identify with the Planck mass, $m_{\rm p}$. One does not however need
a constant of dimension length. It now happens to be the case that in four
dimensions one can send both Newton's constant $G$ and $\hbar$ to
zero, while keeping the ratio $\hbar/G = m^2_{\rm p}$ fixed. This corresponds to 
a limit with $m_{\rm p}$ finite and $l_{\rm p}=0$ that, while not actually being quantum
gravitational, may still capture deviations from SR that
originated in Planck scale effects.

We will look here into an entirely different approach to modified Lorentz-invariance; an approach
that circumvents both the bounds on
Lorentz-invariance violations and the difficulties with locality, and therefore offers
an intriguing new solution to these open problems.  
Knowing that deformations as classical relics 
of quantum gravitational effects have been difficult to reconcile with local field theory, 
we will consider instead a modification that is a pure quantum effect. 
And instead of introducing a fundamental preferred frame on the level of
the action, we introduce a preferred frame only through the process of
measurement. The $\hbar$ is thus instrumental but, at least for the purpose
of this paper, we will not aim to describe gravitational effects and
restrict ourselves to flat space.

This paper is organized as follows. In the next section we introduce
the basic idea and its formalism, and in section \ref{inter} lay out
the physics of interactions. Section \ref{causality} is dedicated to
locality and causality, and in section \ref{cons} 
we discuss some phenomenological consequences, though the details shall
be left for a future work. We discuss assumptions made 
and questions left open 
in section \ref{disc} before concluding in \ref{conc}. 

In the following we use the unit convention $\hbar=G=1$. Small Greek
indices run from $0$ to $3$, and an arrow indicates the spatial
component of a four-vector, e.g. $\vec a = (a_1, a_2, a_3)$. The signature of the 
metric is $(1,-1,-1,-1)$. The constant $c_*$ denotes the usual speed of light.

\section{Quantum mechanics with superpositions of the speed of light}
\label{setup}

Let space-time be a four dimensional manifold ${\cal{M}}$ with Lorentzian signature. We equip it with quantum properties by 
a metric that, instead of being a tensor valued function on ${\cal{M}}$, is an
operator $\hat {\bf g}$ that acts on a wavefunction $| {\bf g} \rangle$ which describes the background spacetime. 
In the general case this operator could have many eigenvalues and -functions, but for our purpose we
focus on a special case: The case in which space-time is flat but the speed of light, $c$, takes on different
values. Then the metric can be in a superposition of eigenfunctions to different $c$'s, 
described by the corresponding Minkowski metric $\eta_{(c)}^{\mu\nu} = {\rm{diag}}(1, -c^2,-c^2,-c^2)$. 
These eigenfunctions fulfill the equation
\beqn
\hat {\bf g} | {\bf \eta}_{(c)} \rangle = \eta_{(c)} |  \eta_{(c)} \rangle ~, ~ 
| {\bf g} \rangle = \intsum dc ~\alpha(c) |\eta_{(c)} \rangle ~, 
\eeqn
with
\beqn
\langle \eta_{(c')}|\eta_{(c)} \rangle = \delta_{cc'} ~ {\mbox{and}}~
 \intsum dc ~ \alpha(c)^* \alpha(c) = 1 ~,
\eeqn
where the asterisk denotes complex conjugation. The existence of these superpositions of metrics is
natural if one considers them as different lapse functions over the same manifold.

The sum (or integral) should actually be taken for all components of the tensor, but since we are
interested only in different values of the speed of light, we can rewrite it into a sum over $c$. 
For the sake of readability, we will in the following restrict ourselves to the case where
$c$ has a discrete spectrum and the eigenvalues are being summed over. Here and in the
following an index in round brackets, $(c)$, indicates dependence on the speed of light. Its 
position does not matter, and we will thus place this index where it is typographically most convenient. 

Since $c$ is a dimensionful quantity, one can make a coordinate transformation that
rescales the spacelike coordinates and changes the $\eta_{ii}$ to some other value. But there is
no one coordinate transformation that will bring all values of the $\eta_{(c)}$'s to agree. 
One should keep in mind though that physically relevant is not the actual value of one $\eta_{(c)}$ but just the ratio
between two with different $c$'s. Usually, if $c$'s are not set equal to one to begin with, they are rarely put into the
metric. We chose this convention here because it has the merit that all other quantities obtain 
their physical units, i.e. $x_0$ has the dimension of a time, $E$ has the dimension of an energy
and $p$ has the dimension of a momentum. 

We are concerned here with the case in which space-time is homogeneous so that the states $|\eta_{(c)}\rangle$ are constant,
a condition that is an approximation for the case in which gravitational effects are negligible. 
The expectation value of $\hat g_{00}$ we name $c^2_*$.  It is the speed of light which we have 
experimentally measured and the standard deviation 
should be small. Exactly how small is a question of constraints from 
available data. The spectrum of the operator ${\hat {\bf g}}$ should actually be derived
from a theory of quantum gravity. In the absence of such a theory, we treat
it as input for the model that will be described here. Our aim is to parameterize the possible
effects. Of course the restriction to
flat space is a very special case, but it is a good starting point to develop
the idea.

To every $\eta_{(c)}$ there is
a Lorentz-group with transformations $\Lambda_{(c)}$ that keep $\eta_{(c)}$ invariant. The operator $\hat {\bf g}$ is invariant under the appropriate
application of the corresponding $\Lambda_{(c)}$ to the subspace spanned by the eigenvector $\eta_{(c)}$. That is, the action of a 
unitary representation $U(\Lambda)$ of a Lorentz transformation $\Lambda$, specified by its group parameters, is given via the eigenfunctions as
\beqn
\hat {\bf g}' |  \eta'_{(c)} \rangle =  U(\Lambda) \hat {\bf g} |  \eta_{(c)} \rangle = \Lambda_{(c)}^T \eta_{(c)} \Lambda_{(c)} | \eta_{(c)}' \rangle = \eta_{(c)} | \eta_{(c)}' \rangle~. 
\eeqn 
Thus, $| \eta_{(c)}' \rangle = | \eta_{(c)} \rangle$ for each eigenvector, and the background in a superposition
state remains invariant $| {\bf g} \rangle = |{\bf g}'\rangle$. 

The above equality is
fulfilled even if the generators are different for different values of $c$. This means
that in principle we can perform a different Lorentz-transformation on each subspace. For one distinct transformation,
we have to match the generators for different subspaces suitably together. We do this by choosing the same generators of rotations
for each subspace, and for the boosts we rescale the velocity so that, in the subspace belonging to $c$, the
velocity $v$ that parameterizes the boost is related to the velocity $v_*$ in the $c_*$-subspace
by $v_*/c_* = v/c$. In other words, we match the transformations so that the commonly used 
quantities $\beta = v/c$  and $\gamma^{-1} = \sqrt{1-\beta^2}$ remain the same in all subspaces. 
 
The transformations $\Lambda_{(c)}$ for different values of $c$ are equivalent representations of the Lorentz-group, i.e. there
exists a matrix $S$ that fulfills
\beqn
S \Lambda_{(c)} S^{-1} = \Lambda_{(c')} 
\eeqn
for each $c, c'$. The matrix $S$ is diag$(c'/c, 1, 1, 1)$. The Lorentz-symmetry that we invoke here thus
is not new in the sense that we use only the well-known representations and actions of the Lorentz-group. Normally however,
the parameter $c$ is considered fixed by experiment to one particular value, and the other, equivalent, Lorentz-transformations
are not regarded as physically interesting. The novel idea here is to relax this restriction on the value of $c$ and
consider the whole set of transformations. 

While this symmetry transformation combines the symmetries of all the subspaces, it is not particularly
useful for maintaining a space-time picture. That is because with these transformations a change of coordinates is
performed differently in each $c$-subspace so that one obtains a whole set of coordinates. If we chose
the coordinates to be the same for each $c$-subspace in one frame $\Sigma$, then a transformation to a frame $\Sigma'$,
moving with some relative velocity, would have different results depending on $c$. Keeping in mind that the 
coordinates are the ones constructed with Einstein's synchronization
procedure (sending light signals back and forth), this is what one expects: The coordinates no longer agree after a boost
because different subspaces use different speeds of light for synchronization.  And while this 
transformation behavior is what one gets if one has an observer for every $c$, each of
which has a different notion of simultaneity, we want instead to restrict 
ourselves to observers in the $c_*$-subspace and transform coordinates according to his, i.e. our normal, $c_*$-transformations. 
The consequence is then that the transformation behavior of the elements of the other subspaces, and functions defined on these, has
to be adjusted.

To see how this works, let us leave aside quantum mechanics for a moment, and consider a 
classical particle with (constant) momentum ${\bf p}$ moving on a trajectory ${{\bf y}}(\tau)$ with 
tangential vector $t_\nu$. We endow the particle with a modified transformation behavior by requiring that
$f(t) = p^{(c)}_\nu t^\nu = 0$ is fulfilled in all reference frames and $f(t)$ transforms like a scalar function from $\Sigma$ to $\Sigma'$, so
that $f'(t') = f(t)$. The momentum $p^{(c)}_\nu$ in addition be lightlike with respect to
$c$, i.e. $p^{(c)}_\nu p^{(c)}_\kappa \eta_{(c)}^{\nu \kappa} = 0$ in all frames. 

The momentum of the particle then transforms under $\Lambda_{(c)}$ and
${\bf y}$ under its inverse. The interpretation of this is clear so long as we are talking about a curve: It is just a curve with an unusual
transformation behavior. But now instead of a curve let us consider a scalar field whose value we want to know for
different $x_\nu$, where $x_\nu$ are our usual coordinates transformed with $\Lambda_{(c_*)}$. Then
we have to adjust the transformation behavior of the defining equation, such that $f'(x') = f((\Lambda_{(c)})^{-1} \Lambda_{c_*} x) $. 
In $\Sigma'$ is then correctly $f'(x') = p^{(c)'}_\nu x'^\nu = p \Lambda_{(c)})^{-1} \Lambda_{(c_*)} x$. The additional factor in the transformation
of the scalar function ensures that the contraction of the vector ${\bf x}$ that transforms differently than ${\bf p}$ remains
invariant. Now, to return to quantum mechanics, if $|\psi\rangle$ is a state of a scalar field, then $\langle \eta_{c}| \psi \rangle = f(x)$ 
is a scalar function which has to obey the same transformation behavior.

At this point, it is useful to note that the partial derivative $\partial/\partial x^\nu$ of $f$ transforms under
$\Lambda_{(c)}$, i.e. it transforms like the momentum. If one wants to evaluate a $c$-momentum in the $c_*$-background,
one takes $p^\nu_{(c)} p^{\kappa}_{(c)} \eta^{(c*)}_{\nu\kappa}$. Since this contraction, while well-defined, is not invariant, 
the location of the indices matters. We take the contravariant momentum vector because otherwise the momentum would
not be parallel to the tangential vector of the curve $y$, which does not make physical sense. A lightlike $c$-momentum in the
$c_*$-background then appears spacelike iff $c>c_*$ and timelike iff $c<c_*$, as one expects. 

We usually do not measure $g_{00}$, except possibly for gravitational waves, which however so far have not been directly detected. 
Instead, we measure the speed of particles in some spacetime background. We are therefore interested in a composite
system
\beqn
|\Phi \rangle = | g , \Psi \rangle ~,
\eeqn 
where $|\Phi\rangle$ is the complete wavefunction including the background and $ \Psi$ describes a particle in that background. 
If the particle is characterized by quantum numbers collectively called $q,c$ with eigenstates $| q, c\rangle$, its expansion is 
\beqn
|\Phi \rangle 
&=& \sum_c \int dq~ \alpha(c) |  \eta_{(c)}  \rangle | q , c  \rangle ~.
\eeqn 
In a more general case, $\alpha$ could be a function also of $q$, but for now we will not consider this
dependence. We will come back to this possibility in the discussion.

To address the question of the invariance of the speed of light, we want to describe the 
propagation of the wavefunction. To that end, we
start with the Klein-Gordon equation for a massless particle in the position representation, such that the quantum numbers
$q$ are the particle's three momenta $\vec p$.
We generalize the box operator so that it takes into account that the metric is now also an operator
\beqn
\quabla = \partial_\mu \partial_\nu \hat g^{\mu \nu}~.
\eeqn
It is then
\beqn
\quabla |\Phi \rangle &=& \sum_c \int d^3p~ \alpha(c) \partial_\mu \partial_\nu \hat g^{\mu \nu} |  \eta_{(c)}  \rangle | \vec p, c \rangle \nonumber \\
&=& \sum_c \int d^3p~ \alpha(c) \partial_\mu \partial_\nu \eta_{(c)}^{\mu \nu} |  \eta_{(c)}  \rangle | \vec p , c \rangle~.
\eeqn
This expansion will fulfill the Klein-Gordon equation when 
\beqn
|\vec p , c\rangle =: v_{\vec p, c} (x) = \propto e^{- i (Et - \vec p \cdot \vec x)} ~ {\mbox{with}} ~ \delta(E - p c)~, \label{sols} 
\eeqn
where $p = |\vec p|$. For momentum eigenstates, this is the usual solution, but every momentum 
now corresponds to a superposition of different energies, depending on the value of $c$. This becomes clearer if we pick one 
momentum eigenstate $\vec p_*$:
\begin{eqnarray}
|\Phi \rangle = \sum_c \alpha(c) e^{-i (c p_* t - \vec p_* \cdot \vec x)} |  \eta_{(c)}  \rangle ~. \label{ave}
\end{eqnarray}
 We reproduce the standard limit
for $\alpha(c) = \delta_{cc_*}$, in which case the background spacetime is in an eigenstate to $c_*$ and 
we can deal with just the field $| \Psi \rangle$ in that background, where it fulfills the usual Klein-Gordon equation.

The eigenmodes are Lorentz-invariant in the way discussed above for the scalar function. We could either 
render $|\Phi \rangle$ Lorentz-invariant by making the transformations on $x^{\nu} = (t,\vec x)$ $c$-dependent,
then we would get
\begin{eqnarray}
|\Phi' \rangle = \sum_c \int d^3p~ \alpha(c) e^{-i (E' t' - \vec p' \cdot \vec x')} |  \eta_{(c)}  \rangle ~, 
\end{eqnarray}
with ${\bf p}' =  {\bf p} \Lambda_{(c)}$ and ${\bf x}' = \Lambda_{(c)}^{-1} {\bf x}$ and $c$ remains invariant. But, as
previously discussed, that
is not useful because the meaning of these coordinates is ambiguous. Instead, we want to keep coordinates that
transform all as ${\bf x}' = \Lambda_{(c_*)}^{-1} {\bf x}$, and then we transform the eigenmodes 
as $v'_{p'}({\bf x}') = v_{p}((\Lambda_{(c)})^{-1} \Lambda_{c_*} {\bf x})$.

To give a mass to the scalar field, one uses the operator $\quabla - c_*^4 m_*^2 \hat f $, where $m_*$ is the (measured)
mass of the particle in the $c_*$-background and $\hat f |\eta_{(c)} \rangle$
may return any dimensionless function of $c$ and $m_*$ as eigenvalues to eigenvectors $| \eta_{(c)} \rangle$. 
While $f(c) = c^4/c_*^4$ suggests itself, there is a priori no obvious
relation between the masses in the different $c$-subspaces, because, for what the symmetry is concerned,
not only $E^2 - c^2 p^2$ is invariant and of the proper dimensionality, but so is its product with any 
dimensionless function of $c$. Thus, without more insights into the mechanism of mass generation,
we have to treat the particle's mass in another $c$-subspace as a parameter of the model. In the following,
to slim down notation, we will write $m^2 = m_*^2 \hat f $ and keep in mind that $m$ is an operator and
its value in some $c$-subspace not necessarily
the measured mass of the particle.

One proceeds similarly for spinors. First, one generalizes the $\gamma$-matrices to
\begin{eqnarray}
\{ \gamma^I, \gamma^J \} = \eta^{IJ} ~, ~ \gamma_{(c)}^\nu = e^\nu_{(c) I} \gamma^{I} ~,
\end{eqnarray}
where quantities with capital Latin indices have the speed of light normalized to one,
i.e. $\eta^{IJ} = {\rm{diag}}(1,-1,-1,-1)$, and 
\begin{eqnarray}
~ e^\nu_{(c) I} = {\rm{diag}}(1,c,c,c)~, 
~  \eta_{(c)}^{\nu\kappa} = e^\nu_{(c) I} e^\kappa_{(c) J} \eta^{IJ}~.
\end{eqnarray}
Then, in the Dirac equation, one replaces the $\gamma^\nu_{c_*}$ with $\hat \gamma^\nu$ that has eigenstates with the property
\begin{eqnarray}
\hat \gamma^\nu |\gamma_{(c)} \rangle = \gamma^\nu_{(c)} |\gamma_{(c)} \rangle ~,
\end{eqnarray}
to obtain
\begin{eqnarray}
(i \hat \gamma^\nu \partial_\nu - c^2_* m) |\Phi\rangle = 0 ~.
\end{eqnarray}
As in the scalar case, the solution to this equation is a superposition of the solutions to the Dirac-equation
for the subspaces of the eigenvectors, i.e. different values of $c$.  

One can use the $e^\nu_{(c) I}$ to convert the transformation behavior of tensors (similar to the way one uses the vierbein to convert from
coordinate transformation to a local transformation behavior). If $V^\nu$ transforms
under $\Lambda_{(c)}$, then $e^\nu_{I (c')} e^I_{\mu (c)} V^\mu$ transforms under $\Lambda_{(c)}$. One can use this
to define a new momentum for the $c$-particles $\tilde p^\nu = e^\nu_{I (c^*)} e^I_{\mu (c)} p^\mu$ that transforms like a normal
Lorentz-vector. However, one then gets a factor $c/c_*$ in the phase of wavefunctions, and the wave-velocity
is not given by $\tilde p_0 / | \tilde p |$. This redefinition thus is not very physical and we will not use
it, though one could work instead with this quantity and carry around the factors.

So far, we have considered only the evolution of the quantum state, now we will look at
the measurement. We will assume that the process of measurement produces an observable
that henceforth transforms under the $c_*$-representations. The process of measurement
also picks out one particular restframe that plays the r\^ole of a preferred frame once
the measurement has been made. One may add this additional r\^ole of the measurement as an axiom to
the standard interpretation of quantum mechanics, but it comes about naturally if the
collapse is replaced by environmentally induced decoherence, where the environment selects
the frame. 

We will thus assume that a measurement of observable $\hat O$, represented by a
hermitian operator, performed in a frame $\Sigma$ collapses the state to 
\begin{eqnarray}
|\Phi \rangle \longrightarrow | O \rangle_\Sigma ~,
\end{eqnarray}
where $|O\rangle$ is an eigenstate of $\hat O$ and the index $\Sigma$ means it is expressed in
the measurement's frame. If decoherence is induced by entanglement with a thermal bath,
as it is typically assumed, then the frame is the restframe of the bath. 

The probability for the measurement outcome is as usual.
The additional assumption here is that 
the measurement outcome $O$ is a classical quantity (a number on an LCD screen) that transforms 
under the Lorentz group $\Lambda_{c_*}$. This means that the measurement reduces the extended Lorentz-symmetry
with the additional parameter $c$ to the usual one along with the transition from quantum
to classical. 

The expectation value of the momentum operator $\hat P_\nu = i \partial_\nu$ is
\begin{eqnarray}
\langle \Phi | \vec P | \Phi \rangle_\Sigma &=& 
\sum_{c,c'} \alpha(c) \alpha(c') \vec p_* \langle \eta_{(c')} |\eta_{(c)} \rangle 
= \vec p_*,
\end{eqnarray}
which is indeed the momentum of the momentum eigenstate. If one prefers to use the $x_{(c)}^\nu$, one uses the operator
\begin{eqnarray}
P_\nu = i \sum_c \partial^{(c)}_\nu ~ {\mbox{with}}~ \partial^{(c)}_\nu = \frac{\partial}{\partial x_{(c)}^\nu}  ~.
\end{eqnarray} 
For the energy $E=P_0$ one has similarly
\begin{eqnarray}
\langle \Phi | E | \Phi \rangle_\Sigma 
&=& p_*  \sum_c c \alpha(c)\alpha^*(c) = p_* c_* ~.
\end{eqnarray}
If we do the transformation into a
different restframe $\Sigma'$ with relative velocity $v$ before measurement, we obtain according to the above
for each energy value in the sum
\beqn
E_*' &=& \frac{1}{\sqrt{1- (v^2/c^2) (c/c_*)^2}} (E_* - v \frac{c}{c_*} p_* ) \nonumber\\
&=&  \sqrt{\frac{v-c_*}{v+c_*}} E_* ~,
\label{red}
\eeqn
which is just the usual relativistic Doppler redshift! In particular it does not depend on $c$ and can be pulled
out of the sum.  Since $c$ is
invariant, this means 
in that case it does not matter in which reference frame one calculates the expectation value, and one can
omit the index $\Sigma$ since it transforms under the usual SR transformation anyway. 

But note that even for $\alpha(c,\vec p) = \alpha(c)$ the eigenvalue of a single measurement, if $c \neq c^*$, does 
no longer transform the same way before and after measurement. Consider we have measured 
the specific value $\tilde c$ with probability $\alpha^2(\tilde c)$.
Then, a Lorentz-transformation after measurement gives
\begin{eqnarray}
E_*' = \frac{1 - v/\tilde c}{\sqrt{1- v^2/c_*^2}} E_*  ~,
\end{eqnarray} 
whereas a Lorentz-transformation before measurement would have resulted in (\ref{red}).
In particular, the velocity $\tilde c$ itself transforms after measurement under the usual 
addition law and is no longer invariant.

To summarize this section, we have seen that we can extend Lorentz-symmetry so that it
accommodates different invariant speeds of light, so long as the state is in a quantum superposition.
We assumed that the process of measurement does not only reduce the
superposition to an eigenstate, but does at the same time reduce the symmetry to the normal
Lorentz-symmetry. As a result the probability distribution over different values of
the speed of light is invariant, but the outcome of any one measurement no longer is.

\section{Quantum field theory and interactions}
\label{inter}

With these prerequisites from quantum mechanics, we can now look at the 2nd quantization. 
We expand the field as
\begin{eqnarray}
\phi(x) = \sum_c \int d^3 p ~ \hat a_{c,\vec p} v_{c,\vec p}(x) + \hat a^\dag_{c,\vec p} v^*_{c, \vec p}(x)  ~,
\end{eqnarray}
where the $v_{c,\vec p}(x)$ are the solutions (\ref{sols}) to the free particle wave equation, and
\begin{eqnarray}
a_{p,c} | \eta_{(c')} \rangle = 0 ~,~ a^\dag_{\vec p,c} |\eta_{c'}\rangle =  \delta_{cc'} |\vec p, c \rangle |\eta_{(c')} \rangle~, 
\end{eqnarray}
and repeated action of creation operators produce multi-particle states in the $c$-background.
For a scalar field the annihilation and creation operators fulfill the commutation relation
\begin{eqnarray}
[a_{\vec p,c}, a^\dag_{\vec p',c'}] = \delta_{cc'} \delta(\vec p - \vec p') ~.
\end{eqnarray}
For spinor fields, one takes the appropriate spinor coefficient functions and anticommutation relations.

To proceed, we now have to investigate which products of fields we can construct invariantly in order to find out
which interaction terms are allowed. For the gauge
fields, $A_\nu$, we use the Lagrangian
\begin{eqnarray}
{\cal L}_{\rm{g}} = - \frac{e^2}{4} \hat g^{\mu \kappa}F_{\mu \nu} F_{\kappa \gamma}~,
\end{eqnarray}
where $F$ is the field strength tensor as usual, $e$ is the coupling constant, and the $c$-value of the fields is 
determined by the $c$-subspace of $\hat {\bf g}$.
This means that four-boson vertices in the non-abelian case cannot mix different $c$-values.

In the Lagrangian for fermions
\begin{eqnarray}
{\cal L}_{\rm{f}} = \overline \psi (i \hat \gamma^{\nu} \partial_\nu - m ) \psi  + cc. ~,
\end{eqnarray}
the $c$-value of the fermion is determined by the $c$-subspace of $\hat \gamma$. The interaction term
takes the form
\begin{eqnarray}
{\cal L}_{\rm{int}} = e M \overline \psi \hat \gamma^{\nu} A_\nu  \psi ~,
\end{eqnarray}
with the transition matrix
\begin{eqnarray}
\langle \eta_{(c)} | M | \eta_{(c')} \rangle = M_{cc'} ~,
\end{eqnarray}
and $M=M^\dag$. $M$ is not necessarily diagonal because the $c$-subspace of the fermions does not 
need to be the same as that of the gauge field. This is
because, as noted earlier, we can put in a factor $e_{(c')} e_{(c)}$ to adjust the transformation behavior of
the $\gamma$'s (that are contracted with the partial derivative acting on the $c$-spinor and produce
a $c$-momentum) to that of the $A_\nu$. Or, in other words, the relevant property characterizing
the symmetry of the gauge field is the transformation of the phase and not the transformation of the
polarization vector. $M$ is a matrix that, when projected on the $c$-eigenspaces, 
encodes the coupling between different $c$-sectors. These vertices then mix different $c$-values of 
fermions and gauge bosons. 

If one inserts the field expansion in such a Lagrangian, the Feynman rules in momentum space are then the normal 
ones with the following additions: 
\begin{itemize}
\item Every vertex obtains a factor $M_{cc'}$, one $c$ for the fermions, one for the gauge boson.
\item The external ingoing lines belong to the same $c$-subspace. External outgoing lines also belong to the same $c$-subspace,
but not necessarily the same as the ingoing ones.
\item Vertex indices must be matched to the coupling particles' transformation behavior. 
\item Sum over all $c$'s of virtual particles.
\end{itemize}
And, as laid out in the previous section, the momenta transform under the respective $\Lambda_{(c)}$ until measurement,
after which they become a $c_*$ four vector. 
 
The amplitudes then have as usual a $\delta$-function for conservation of the four-momentum. Note
that the argument of this $\delta$-function ${\it differs}$ from one frame to the next. It
does not differ by a coordinate transformation, it is actually a different argument.
It is only the measurement that one selects one. As we have seen in the previous section, 
the outcome depends on
the frame of the measurement and disagreements are unobservable. 

\section{Locality and Causality}
\label{causality}

Locality can become a problematic concept in theories in which the speed of light 
can take different values but still observer independence
should be fulfilled. The reason is that with the requirement that different
values of speeds remain invariant, space-time points have no
well-defined transformation behavior: The location of a point after a change of 
reference frame depends on which speed is kept invariant. In particular, 
a point defined in one reference frame by different means though intersecting
curves can, after a change of reference frame, split up into various points.
For the case of {\sc DSR}
this has been shown in \cite{Hossenfelder:2009mu,Hossenfelder:2010tm}. Recent
suggestions for how to address the problem have built up to a new `Principle of
Relative Locality' \cite{Smolin:2010xa,Jacob:2010vr,AmelinoCamelia:2010qv,Smolin:2010mx,AmelinoCamelia:2011bm}.
This approach accepts the arising nonlocality and aims to show it is not
problematic after all. (For some discussion, see also \cite{Hossenfelder:2010yp,Hossenfelder:2010jn,Hossenfelder:2010xr}.)

We too encountered in the previous section the need to transform coordinates 
depending on the value of $c$, reflected in the set of transformations $\Lambda_{(c)}$. But our 
approach offers an entirely new solution for the problem. The observer-independence of the 
speed of light is now a fundamental property of the evolution of a quantum state, but 
each single measurement outcome depends on the restframe in which the measurement was made. 
In particular, a speed of photons different from the average value $c_*$ will after measurement 
no longer be an invariant of the transformation, but transform under normal Lorentz-transformations $\Lambda_{(c_*)}$. 
Thus, disagreements in different observers' definitions of a point due to different 
invariant speeds are never reflected in actual observables. 

The `Box-problem' discussed in
reference \cite{Hossenfelder:2009mu,Hossenfelder:2010tm} is circumvented because observers never
disagree on the outcome of the measurement. In {\sc DSR} the particle's worldline transforms
under a non-standard Lorentz-transformation that is energy-dependent. As a consequence, the
statement whether three lines meet or do not meet in one point depends on the reference frame,
and (to some precision) the question whether they meet is a requirement for local interactions to
take place. In the
scenario discussed here, in contrast, making the measurement in one frame fixes the eigenvalue of the
speed and a different observer would interpret the speed to be the {\sl normal} Lorentz
transformation of the speed measured. The measurement
is either made in the laboratory frame, in which case the bomb blows up and the observer in the 
satellite agrees, or it is made in the satellite frame, in which case the bomb does not blow up
and the observer in the lab agrees. The situation is entirely symmetric as long as both frames
represent identical measuring processes with some relative velocity.

Solving the problem with locality does however not solve the problems with causality that
superluminal information exchange creates. Indeed, it seems one has to give one up for
the other. One creates a problem with locality if
there exist curves with different transformation behaviors because their intersections
can be used to define points. If one does not accept these non-localities, the need to 
reproduce SR in the limit of non-quantum objects means that everything
that we can plausibly refer to as an observer transforms under normal Lorentz-transformations.
But a normal Lorentz-transformation can turn a superluminal curve into one going backwards
in time. In {\sc DSR} on the other hand, the modified transformation behavior
of the speed-of-light allows it to remain in the upper, $t>0$, part of {\it each} reference frame.

To be more precise, the problem with causality is not that a curve that in one 
reference frame is superluminal seems to be going backwards in time in another frame, because 
the curve itself does not have a direction. Both observers could interpret the particle as moving forward
in time but into opposite spatial directions. Closed curves in flat Minkowski-space then are not a
problem fundamentally; one just has to demand consistency. This means if a particle at
some time $t_0$ could affect its own earlier curve at $t_1 < t_0$, this would have been taken into
account at the time $t_0$ already. 

\begin{figure}[ht]
\includegraphics[width=7.0cm]{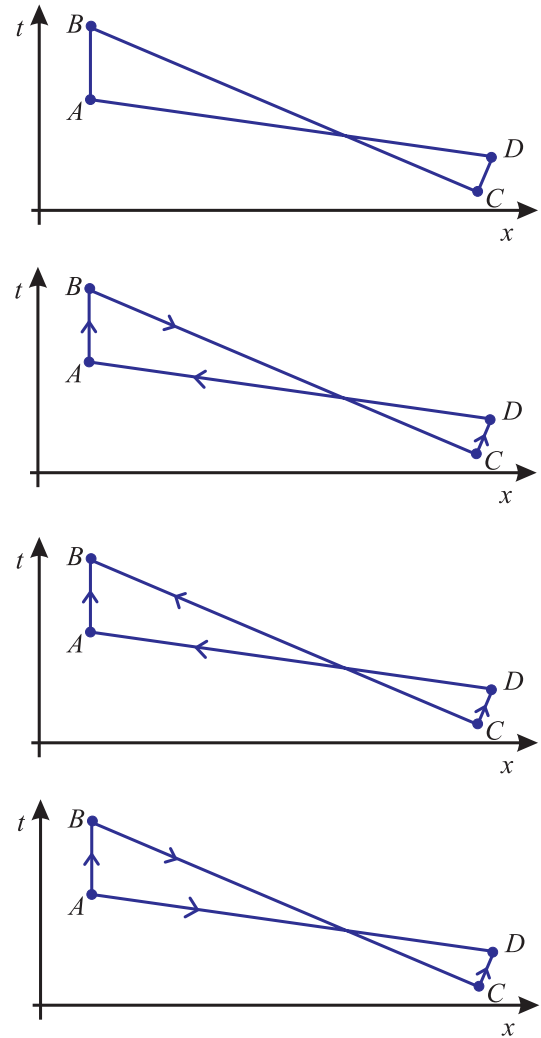}
\caption{{\small Closed curves. Top: Without arrow of time. 2nd from top: With inconsistent arrows of time, generating
the possibility for grandfather paradoxa. 2nd from bottom and bottom: With consistent arrow of time. Closed curves could be constructed with
only 3 straight lines. We have included a fourth to show that taking into account the finite amount of time necessary to process
information does not remove the problem.}}
\label{4}
\end{figure}

As an example, consider the closed curve in Figure \ref{4}, top,
and interpret each corner as a scattering event. One could read this curve as a particle that
propagates freely from $A$ to $B$, scatters in $B$ and produces a particle with superluminal momentum
that then moves towards $C$ which, in the chosen reference frame is earlier in time than $B$. In $C$ and in $D$ 
the particle scatters again, produces an outgoing particle that then intersects with the original particle's previous
curve in $A$. Now on the level of elementary particles, this would just mean that the state of the
particle at $B$ would have had to take into account event $A$ already, because it was always there.  
We can also read the curve the other way round which (in the chosen example) would correspond
to some particle going from $C$ to $D$, scattering twice and creating to superluminal
particles going to $A$ and $C$ respectively.

However, this does no longer work once we take into account that, for better or worse, our
world is evidently not time-reversal invariant and we do have an arrow of time that points 
somewhere we can for lack of a better word call `forward.' Problems with causality can no longer simply be solved by 
demanding consistency in this simple form when we consider macroscopic objects that display an arrow of time because 
it becomes possible to create paradoxa (that have
been extensively used and abused in the science-fiction literature). 

To see the difference,
consider in $A$ you fill out a lottery ticket. On the way to $B$ you learn that you didn't win
but write down the winning numbers and send them to your friend at $C$. Your friend then,
at $D$, sends the numbers back to your earlier self at $A$. Consistency would now demand that you
already knew the numbers all the time, in which case you would have won the lottery
already and still there was no avoiding that you
will send the numbers to your friend. Alternatively, your friend is not able to send you the
numbers, or you will not be able to receive them. 
But apart from that being what consistency demands, it is difficult to see exactly
what cosmic conspiracy would prevent you or your friend from sending the numbers back and forth and creating
a paradox. 

The difference to the case of elementary particles is that this story has a direction
of information flow and a notion of `learning'. The process of `sending' is different to the 
process of `receiving' because the sender previously knew of the information whereas the 
receiver does not. If we time-reverse this process it looks very different. (One does not
untype an email when one receives it.) For macroscopic
objects thus the curve would have to be endowed with arrows indicating a direction, depicted
in Fig. \ref{4}, 2nd from top, that inevitably have to run backwards in time somewhere (in any reference frame).
That is what creates the problem.

But this analysis of the problem also contains the seed for its resolution.
Since there is no point denying the existence of an arrow of time, we have to take
it into account consistently. Recall that in our framework it is only after measurement that
the curve of a superluminal particle is subject to SR transformations. All
we have to do is to endow the process of measurement with an arrow of time
which comes naturally through the framework of environmentally induced
decoherence. Thus, there is an environment that creates an arrow of time,
which is a vector field timelike in the $c_*$ frame, that tells us which
end of a curve is the emission and which is the detection. It is then
impossible to send information backwards in time in the environment's
restframe, and while it is still possible to send it backwards in time
in some other frame, it is no longer possible to send information in
a closed curve. Instead, the arrows of information flow all have to point 
forward in some frame, resulting in the situation depicted in Fig. \ref{4}, two
bottom figures (depending on how the depicted $t$ relates to the arrow of time). 
Concretely, this means you in $B$ cannot prepare a state that your friend
can measures in $C$, or your friend in $D$ cannot prepare a state that you
measure in $A$. 

This of course does introduce a preferred frame. But note that this
preferred frame, one we know exists, arises through interaction with the environment 
in the measurement processes and is not present through a {\sc LIV} term in the Lagrangian. 

Circumventing of timelike closed curves by requiring
consistency of history has become known as ``Novikov's self-consistency principle'' \cite{Friedman:1990xc}. 
It should be noted that the simple way that timelike closed loops are prevented here
is possible only because we are still dealing with a flat, topologically trivial, 
background. The timelike closed curves that occur here cannot be geodetic and still intersect 
with themselves; they necessitate a change of direction, and thus an interaction, in at least
three points. The reader might be familiar with the self-consistency principle from the
general relativistic case, where the issue of timelike closed loops is much more 
involved, and a solution like the one discussed above is in general not possible, or
at least not physically justified. For some discussion see also \cite{Visser:2002ua}.

It may be of interest to the reader that Geroch \cite{Geroch:2010da} has argued on
very general grounds that the existence of different causal cones is possible without
being in conflict with SR. In Geroch's work it was however not investigated the transformation
behavior of these cones and their invariance in particular. 

\section{Possible phenomenological consequences}
\label{cons}

The standard model, and quantum electrodynamics (QED) in particular, are extremely well
tested theories. This raises the question how tightly existing experiments constrain the
possibility discussed here. While we will leave details of the phenomenology to be studied in a future work, 
here we want to discuss some general properties. 

Since we have never noticed a photon moving with anything
else than $c_*$ (to some precision) we should expect the probability for photons
to propagate in other $c$-subspaces to be small, $\ll 1$. Let us consider a very simplified case, 
that in which there is only one other value $c_1$ in addition to $c_*$. With the
convention that $M_{c_* c_*} = 1$ (if not, the factor can be absorbed in the coupling
constant), we have two remaining free parameters: $M_{c_* c_1} = \lambda \ll 1$
and $M_{c_1 c_1} = \mu$ in addition to the masses of the $c_1$-fermions which, as
we noted earlier, stand in no obvious relation to their masses in the $c_*$-subspace.

There can be no modification to Compton scattering,
because both photons and fermions are in the ingoing state. The amplitude for 
Bhabha scattering between $c_*$-electrons obtains an additional contribution 
of order $\lambda^2$ and the amplitude for $c_1$-electrons in the outgoing state
has a factor $\mu \lambda$. Note however that the same factor comes in again
through the amplitude for the detection cross-section. That is to say, if the 
particles are difficult to produce, they are also difficult to detect. 

A not
detected $c_1$-photon or electron would result in missing energy and momentum
that does not fulfill the condition $E^2 - c_*^2 p^2 = m_*^2 c^4_*$. Instead,
it would appear to have an apparent mass $m_{\rm{app}}$ of
\begin{eqnarray}
m_{\rm{app}} = m \sqrt{1 + \frac{p^2 c_1^2}{m^2 c_*^4} \left( 1- \frac{c_1^2}{c_*^2} \right)^2} ~,
\end{eqnarray}
where $p^2$ is the (square of the) three-momentum in the measurement frame. Note that this apparent 
mass can be imaginary if $c_1>c_*$. This apparent mass is the mass that we would assign to the
particle within special relativity. The particle's actual mass, which appears in the Lagrangian, remains
positive and real valued.

However, the best experimental tests of {\sc QED} come from virtual processes that will 
yield constraints on the parameters of this model. The question is whether parameters that 
fulfill the constraints render the model uninteresting already, which will require further study.

\section{Discussion}
\label{disc}

Let us summarize the assumptions made here and discuss their relevance. 

First,
we have assumed that $c$ takes on discrete values. While the formalism presented here
can easily be extended to continuous $c$-values, it seems more plausible that, if there
exist indeed superpositions of different $c$'s, the values are discrete, at least
in the vicinity of $c_*$. If they are not, one would expect the probability to be
smooth in the vicinity of $c_*$. Then, in the absence of a gap in the 
spectrum, one loses the rationale to use only one particular $c$-value, namely $c_*$
for the measurement outcome.

We have further, in section \ref{setup}, considered a
product state between the particle in some background and the background. The more
general case would be that $\alpha$ is a function also of the quantum numbers of the
particle $\alpha(c,q)$. For one, this would be the outcome of some scattering process
in which case $\alpha$ would generically depend on all the momenta of
scattering particles. But one may also consider the possibility that $\alpha(c,q)$ is an 
intrinsic property of the particle. 

If $\alpha(c,q)$ is an invariant and intrinsic property of the particle, then it may be
a function of the momentum. This allows us to relate the here proposed model to {\sc DSR}.
For a photon, the only 
way to obtain a non-trivial
(on shell) momentum-dependence that is also Lorentz-invariant is to use a modified version of
transformations in momentum space.  Then, the speed of light obtained from the expectation values
of energy and momentum may become a function of the particle's energy that in the low energy
limit reproduces normal Lorentz-symmetry, thereby connecting the here proposed model 
to {\sc DSR}.  However, in this
case $c$ would need to have a continuous spectrum (since the energy can be continuously
redshifted). Such a version of the model would still be different from {\sc DSR} in that
the different speeds of light can exist only in superpositions.

\section{Conclusion}
\label{conc}
We have shown here that, next to Lorentz-invariance breaking and deformations of special relativity,
there exists a novel third way how departures from Lorentz-invariance that may arise in
quantum gravitational effects can make themselves noticeable. 

The departure from Lorentz-invariance proposed here arises from
superpositions of different metrics on the same background manifold, so
that for each of the metrics the maximally possible and invariant speed of massless 
particles takes on a different value.
The process of measurement produces observables that obey the laws of special relativity.
To preserve causality, we have assumed that the measurement does introduce a preferred
frame. This modification of Lorentz-invariance makes it unnecessary to introducing 
Lorentz-invariance violating operators,  and does not create problems with locality or 
causality; at least in flat space. 

The here proposed model has phenomenological consequences for particle physics 
that need to be further explored to find out how tightly the parameters of the 
model are constrained by available 
data already. 

\bigskip

\section*{Acknowledgements}

I thank Ben Koch, Jakub Mielczarek, Stefan Scherer and Lee Smolin for helpful feedback and discussions,
and Sean Carroll for drawing my attention to reference \cite{Geroch:2010da}.

\end{document}